# Hydrostatic pressure effect on the pseudogap in $Y_{0.77}Pr_{0.23}Ba_2Cu_3O_{7-\delta}$ single crystals FREE

A. L. Solovjov; L. V. Bludova; A. S. Kolesnik; E. V. Petrenko; M. V. Shytov; A. Sedda; E. Lähderanta; D. Sergeyev; R. V. Vovk

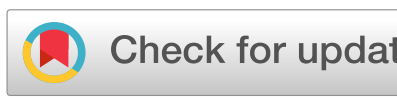

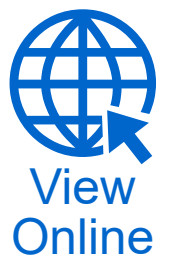

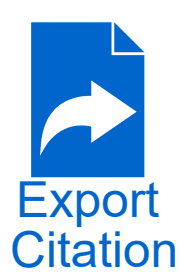

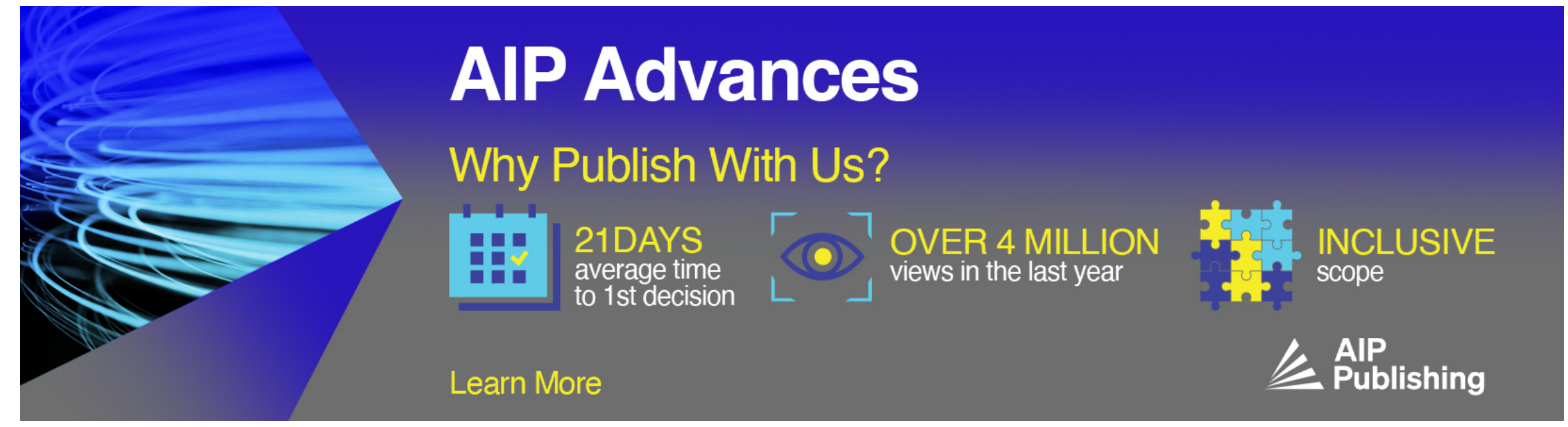

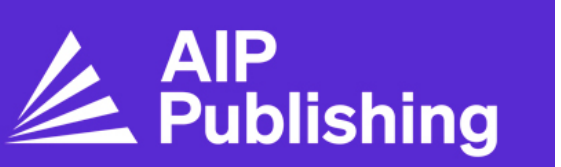

# Hydrostatic pressure effect on the pseudogap in $Y_{0.77}Pr_{0.23}Ba_2Cu_3O_{7-\delta}$ single crystals

View Online

Export Citation

CrossMark

A. L. Solovjov,[1,2,3,a)] L. V. Bludova,[1] A. S. Kolesnik,[1] E. V. Petrenko,[1] M. V. Shytov,[1] A. Sedda,[2] E. Lähderanta,[2] D. Sergeyev,[4] and R. V. Vovk[5]

## AFFILIATIONS

[1]B. Verkin Institute for Low Temperature Physics and Engineering of the National Academy of Sciences of Ukraine, Kharkiv 61103, Ukraine
[2]Department of Physics, LUT University, Lappeenranta 53850, Finland
[3]Institute for Low Temperatures and Structure Research of PAS, Wroclaw 50-422, Poland
[4]K. Zhubanov Aktobe Regional University, 030000 Aktobe, Kazakhstan
[5]Physics Department, V. N. Karazin Kharkiv National University, Kharkiv 61022, Ukraine

[a)]**E-mail:** solovjov@ilt.kharkov.ua

## ABSTRACT

Significant efforts have been made to elucidate the properties of high-temperature superconductors (HTSCs) using hydrostatic pressure (HP). We studied the changes of resistivity $\rho(T)$, superconduting transition tempetature $T_c$, fluctuation conductivity $\sigma'(T)$, and pseudogap $\Delta^*(T)$ of $Y_{1-x}Pr_xBa_2Cu_3O_{7-\delta}$ (YPrBCO) ($x = 0.23$) single crystal under HP up to ~1.1 GPa. Defects created by magnetic inclusions of $PrBa_2Cu_3O_{7-\delta}$ (PrBCO) were found to play a significant role in the behavior of the sample. The largest decrease in $\rho(T)$ was found depending on HP at a rate of $d \ln \rho(100\,K)/dP = -(29 \pm 0.2)$ %·GPa$^{-1}$. This indicates that the mechanisms of the influence of HP on the $\rho(T)$ and $T_c$ of YPrBCO and $YBa_2Cu_3O_{7-\delta}$ single crystals are different. From the analysis of the pseudogap, it was revealed that the mechanism of the interaction of charge carriers with magnetic PrBCO impurities changes three times with increasing HP. Relatively low HP, up to ~0.5 GPa, promotes the formation of defects caused by PrBCO magnetic inclusions. Starting from 0.6 GPa, the HP neutralizes the influence of magnetic impurities, and the magnetic maximum on $\Delta^*(T)$ disappears. Above ~1.0 GPa, HP aligns the magnetic moments of PrBCO, and the magnetic maximum reappears on $\Delta^*(T)$, more pronounced than at $P = 0$. Comparison with the Peters–Bauer theory showed that the local pair density in HTSC does indeed increase with increasing pressure.

## 1. INTRODUCTION

The pseudogap (PG) state, which is opened in cuprate high-temperature superconductors (HTSCs) below the characteristic temperature $T^* >> T_c$, is one of the most mysterious and simultaneously interesting phenomena in modern solid-state physics.[1–3] It is well established that in HTSCs, the PG is observed when the charge carrier concentration, $p$, varies between slightly doped (SD) and optimally doped (OD) levels. Understanding the PG physics would definitely shed more light on the mechanism of superconducting pairing in HTSCs, which is also not fully clarified yet[3] (and references therein). Knowledge of this mechanism is of primary importance for the search for superconductors with an even higher, preferably room temperature, superconducting (SC) transition $T_c$. The temperature dependence of the resistivity of HTSC in the normal state above $T^*$ is always linear. But unexpectedly, it deviates downwards from linearity at $T \leq T^*$.[4,5] It has been convincingly shown that at $T = T^*$, not only does $\rho(T)$ deviate from linearity, but the density of states (DOS) at the Fermi level also begins to decrease noticeably,[6,7] which is just called a pseudogap (PG).[8] Perhaps this is the only experimental result with which the entire superconducting community agrees.

Unfortunately, the explanation for the deviation of $\rho(T)$ from linearity that led to the opening of PG remains highly controversial. A variety of models have been proposed to explain both the pairing mechanism and the PG phenomenon in HTSCs,[1,3,9–13] including even spin–charge separation scenarios,[14] proposed by Anderson. The most modern of these models, which have nothing to do with superconductivity, is the charge density wave (CDW) model, which is recorded in $YBa_2Cu_3O_{7-\delta}$ (YBCO) cuprate by X-ray diffraction[2] (and references therein). However, the maximum amplitude of the CDWs at $p = 0.12$ is approximately 160 K, while the PG opening temperature is $T^* \approx 245$ K. This shows that CDWs also have nothing to do with the PG opening. Also, none of the non-superconducting models explain why $\rho(T)$ deviates downward from linearity at $T = T^*$. Therefore, we share the point of view of another group of researchers that the PG arises as a result of the formation of paired fermions, local pairs (LPs), at $T \le T^*$,[9–12] which is confirmed by many experimental results.[15–20] This approach also explains the deviation of $\rho(T)$ downwards from linearity at $T \le T^*$, since the LPs are considered to transport electric charge without dissipation.[10–12,20] Unfortunately, this approach has not yet become generally accepted and requires additional experimental evidence.

A powerful method for studying HTSCs, which affects the structure, $T_c$, resistivity, fluctuation conductivity (FLC), and PG, is hydrostatic pressure. It also enables validating the adequacy of numerous theoretical models, as well as establishing the most significant parameters of HTSC structures, which determine their physical characteristics in normal and SC states[21] (and references therein). One of the most promising materials for studying PG, especially when applying high pressures, is the YBCO family. This is due to the possibility of variation in its composition by substitution of yttrium (Y) with the isovalent analogues, or changing the degree of oxygen non-stoichiometry. The compounds $Y_{1-x}Pr_xBa_2Cu_3O_{7-\delta}$ (YPrBCO) with partial substitution of Y by praseodymium (Pr) atoms are of particular interest in that respect. The replacement of Y with Pr leads to a noticeable increase in the resistivity ρ and a decrease in the critical temperature $T_c$, which becomes zero at $x \approx 0.7$ (the "praseodymium anomaly").[22–25] It is believed that in the $RBa_2Cu_3O_7$ (R123) compounds, the R–Pr system is an insulator, and the strong hybridization of Pr 4$f$ and O 2$p$ orbitals leads to localization of oxygen holes in the Fehrenbacher–Rice energy zone.[26,27] Thus, PrBCO becomes a dielectric being isostructural to YBCO. The PrBCO dielectric inclusions, arising in the process of manufacturing YPrBCO crystals, form multiple defects in the YBCO superconducting matrix,[23–27] which significantly affect the transport properties of the sample. Therefore, doping of the $Y_{1-x}Pr_xBa_2Cu_3O_{7-\delta}$ with Pr on the one hand leads to a gradual suppression of superconductivity with increasing the Pr concentration $x$, and on the other, it allows to preserve almost constant both the lattice parameters and the oxygen content (7–δ) of the sample under investigation.[25–28] The $Pr^{+3}$ atoms have an intrinsic magnetic moment of $\mu_{Pr} \approx 3.58\mu_B$,[29] which is $m_{eff} \approx 2\mu_B$ in the PrBCO compound.[30] That is why the study of the effect of Pr impurity on the properties of YPrBaCuO single crystals is considered very promising for revealing the mechanisms of the interplay of the superconductivity and magnetism in HTSC, (see, for example, Refs. 31 and 32), and references therein), which is important for the final clarification of the physical nature of both the PG and high-temperature superconductivity in general.[2,10,33–35]

In cuprates, in contrast to the influence of a magnetic field[36–38] or electron irradiation,[20,39–41] the $dT_c/dP$ dependence is mostly positive, while the derivative $d \ln \rho/dT$ is negative and is relatively large.[42–46] However, the data given in studies of the effect of pressure on the $T_c$ of YPrBCO compounds (see, for example, Refs. 21 and 47) are often inconsistent. In the case of single-crystal samples, difficulties may be caused by the appearance in the system of a sufficiently disordered structure of twin boundaries (TBs).[48–50] The TBs being extended to two-dimensional defects can serve as drains for lower dimensionality defects, which in turn represent strong scattering centers for normal and fluctuation charge carriers,[21,50] thereby having a significant impact on charge transfer processes in a particular experimental sample. When pressure is applied, the volume of the unit cell decreases, contributing to the ordering of the system, leading to a decrease in the number of structural defects and a decrease in ρ.[21,45] At any rate, the mechanisms of influence of pressure on both $T_c$ and ρ are not fully understood, as the nature of the transport properties of HTSCs is also not completely clear. The main contribution to the conductivity of cuprates is made by the $CuO_2$ planes, between which there is a relatively weak interplanar interaction. It is assumed that the pressure leads to an increase in the density of charge carriers $n_f$ in the conducting $CuO_2$ planes and, as a result, to a decrease in ρ.[21,45] An increase in $n_f$ under pressure should also lead to an increase in $T_c$, i.e., to a positive $dT_c/dP$ value as observed in experiments.[42–45,51] At the same time, there are only a very limited number of studies devoted to the study of the influence of pressure on the FLC and PG in HTSC cuprates,[21,51,52] including lightly doped with Pr ($x = 0.05$).[53] However, to the best of our knowledge, for YPrBCO with an intermediate Pr content of $0.7 > x > 0$, such studies have not been carried out at all.

To find out how defects produced by magnetic cells of dielectric PrBCO affect the properties of non-magnetic YBCO, in the paper we investigate the effect of hydrostatic pressure up to $P = 1.1$ GPa on resistive characteristics, the excess conductivity σ′(T) and the pseudogap $\Delta^*(T)$ of single crystals of $Y_{1-x}Pr_xBa_2Cu_3O_{7-\delta}$ ($x \approx 0.23$) with nearly stoichiometric oxygen content. The geometry of the transport current $I$ flow was chosen parallel to the TBs ($I \parallel$ TBs), which made it possible to minimize the effects of scattering on the twin boundaries.[50,54]

## 2. EXPERIMENT

The $YBa_2Cu_3O_{7-d}$ single crystals were grown using solution-melt technology.[55] The crystal growth and oxygen saturation regimes were the same as for undoped YBCO crystals. $Y_2O_3$, $BaCO_3$, and CuO compounds were used as initial components for crystal growth. In order to obtain crystals in which Y is partially substituted by Pr ($Y_{1-x}Pr_xBa_2Cu_3O_{7-\delta}$), $Pr_5O_{11}$ was added to the initial charge in the appropriate percentage.[32] Rectangular crystals of about $1 \times 2 \times 7$ mm were selected from the same batch to perform the resistivity measurements. The smallest parameter of the crystal corresponds to the $c$-axis. The experimental geometry was selected so that the transport current vector in the $ab$-plane was parallel to the TBs.[32,50,54] A fully computerized setup utilizing the four-point probe technique with stabilized measuring current of up to 10 mA was used to measure the $ab$-plane resistivity, $\rho_{ab}(T)$.[31] Silver epoxy contacts were glued to the extremities of the crystal in order to produce a uniform current distribution in the central region, where voltage probes in the form of parallel stripes were placed.[50] Contact resistances below 1 Ω were obtained. Temperatures were measured with a Pt sensor having an accuracy of about 1 mK. The hydrostatic pressure was generated inside a

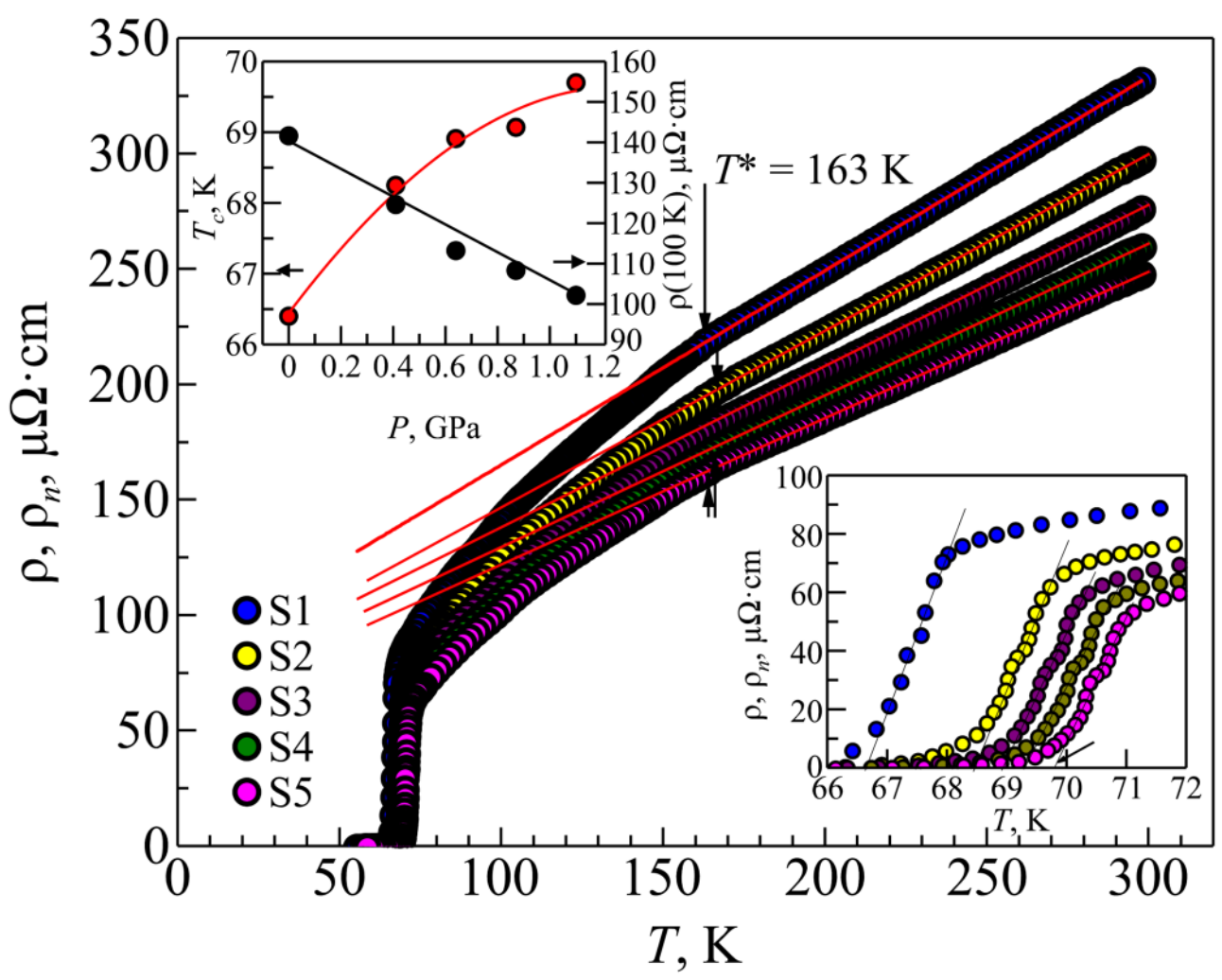

**FIG. 1.** Temperature dependences of the resistivity ρ of $Y_{0.77}Pr_{0.23}Ba_2Cu_3O_{7-\delta}$ single crystal at different $P$, GPa: 0 (S1, blue dots), 0.41 (S2, yellow dots), 0.64 (S3, violet dots), 0.87 (S4, green dots), and 1.1 (S5, magenta dots). Straight solid lines designate extrapolated $\rho_N(T)$. The left inset illustrates the pressure dependence of $T_c$ (red dots) and ρ(100 K) (black dots); the solid curve and line are a guide for the eye. The right inset illustrates how $T_c$ was determined at different pressures.

Teflon cup housed in a copper–beryllium piston-cylinder cell, as described in Ref. 56. A manganin gauge made of a 25 Ω wire was used to determine the applied pressures. Transformer oil was used as the transmitting medium, and pressures were changed at room temperature in the order of increasing magnitude. For each applied pressure, experimental runs were carried out at rates of about 0.1 K/min near $T_c$ and 0.5 K/min at $T >> T_c$.

## 3. RESULTS AND DISCUSSION

### 3.1. Resistance and critical temperature

The set of resistivity curves $\rho(T) = \rho_{ab}(T)$ of the $Y_{0.77}Pr_{0.23}Ba_2Cu_3O_{7-\delta}$ (YPrBCO) single crystal measured at ambient pressure $P = 0$ (blue dots), 0.41 GPa (yellow dots), 0.64 GPa (violet dots), 0.87 GPa (green dots) and at 1.1 GPa (magenta dots), which actually can be considered as 5 different samples (S1–S5, respectively), are shown in Fig. 1. The sample parameters obtained at different $P$ are listed in Tables I and II. Despite the relatively low $T_c$ (= 66.4 K at $P = 0$), at all applied pressures, the dependences ρ($T$) have the shape typical of OD HTSCs.[5,57,58] This result confirms the well-known conclusion that when doping Pr, $T_c$ decreases due to the influence of magnetic impurities PrBCO, leaving the oxygen content unchanged.[23–25] Indeed, in the normal state from $T^*$ to 300 K, the ρ($T$) dependences are linear with the slope $d\rho/dT \approx 0.84$ μΩ cm/K ($P = 0$) and $d\rho/dT \approx 0.637$ μΩ·cm/K ($P = 1.1$ GPa). The slope was determined by computer linear fitting, approximating the experimental dependences ρ($T$) by the straight line equation $\rho(T) = \rho_0 + aT$, where $a = d\rho/dT$ and $\rho_0$ is the residual resistance determined by the intersection of the extrapolated straight line with the $Y$-axis at $T = 0$. The approximation confirmed high linearity of the dependences with a root-mean-square error of 0.002 ± 0.001 in the specified temperature interval for all the samples studied. The linear slope $a$ is found to linearly decrease with $P$ (left inset in Fig. 1) at a rate $da/dP = -(0.185 \pm 0.2)$ μΩcm$K^{-1}GPa^{-1}$, which is about 1.3 times larger than $da/dP = -(0.141 \pm 0.2)$ μΩcm$K^{-1}GPa^{-1}$ obtained for the YPrBCO single crystal with $x = 0.05$.[53] The deviation of ρ($T$) from the linearity shown in the figure by the solid red lines towards lower values determines the PG opening temperature $T^*$. For the sake of a more accurate determination of $T^*$ with an accuracy of ±0.5 K, the criterion $(\rho(T) - \rho_0)/aT = 1$ was used.[59] In this case, $T^*$ is defined as the deviation of $(\rho(T) - \rho_0)/aT$ from 1.[20,21,31,32,59] Both approaches yield the same $T^*$ values. As can be seen in the figure, at $P = 0$, $T^* = 163$ K, and, surprisingly, does not actually change under pressure (Table II). We also note a rather small, compared to pure YBCO single crystals with a low $T_c$,[5,51,60] value of the PG temperature $T^* = (163 \pm 0.3)$ K at $P = 0$. Any alloying substances, including Pr, can play the role of impurities in the sample since their random Coulomb fields create additional scattering centers for charge carriers.[24,25,61] It can be assumed that additional defects induced by PrBCO, as well as the magnetic moment of PrBCO, prevent the establishment of phase coherence and the formation of fluctuating Cooper pairs (FCPs) above $T_c$, consequently reducing $T^*$.[25,32]

At the same time, the increase in pressure leads to a noticeable decrease in the resistance of the sample (black dots in the left inset of Fig. 1). The relative reduction of ρ($T$) as a function of pressure is practically independent of temperature above $T^*$ and amounts to $d \ln \rho(300\ K)/dP = -(26 \pm 0.2)$ %·$GPa^{-1}$. This value is approximately the same as $d \ln \rho(300\ K)/dP = -(25.5 \pm 0.2)$ %·$GPa^{-1}$ obtained for Bi-2212 single crystals,[42] but it is noticeably larger than $d \ln \rho(300\ K)/dP = -(20.6 \pm 0.2)$%·$GPa^{-1}$ obtained for YPrBCO single crystal with $x = 0.05$.[53] Accordingly, $d \ln \rho(100\,K)/dP$ amounts to $-(29 \pm 0.2)$ %·$GPa^{-1}$, which is unexpectedly large compared with $d \ln \rho(100\,K)/dP = -(10.6 \pm 0.2)$ %·$GPa^{-1}$ obtained by us for YPrBCO single crystal with $x = 0.05$.[53] The typical value for YBCO single crystals is $d \ln \rho(100\,K)/dP = -(12 \pm 0.2)$ %·$GPa^{-1}$.[42] Thus, increasing the Pr content obviously enhances the effect of pressure on resistivity. It is believed that high pressure promotes charge transfer (i.e., hole doping of $CuO_2$ planes), which leads to the observed decrease in resistivity.[45] However, strictly speaking, this is somewhat surprising since, as noted above, in YPrBCO some of the charge carriers must be localized, which can suppress the growth of $n_f$ under pressure in the $CuO_2$ planes and lead to a decrease in the intensity of the decrease in $\rho$.

**TABLE I.** Resistive and fluctuation conductivity parameters of $Y_{0.77}Pr_{0.23}Ba_2Cu_3O_{7-\delta}$ single crystal at different pressure.

| $P$, GPa | ρ (300 K), μΩcm | ρ (100 K), μΩcm | $T_c$, K | $T_c$ , K | $T_0$, K | $T_{01}$, K | $T_G$, K | $d_{01}$, Å | $\xi_c(0)$, Å |
|---|---|---|---|---|---|---|---|---|---|
| 0 | 331.24 | 141.16 | 66.4 | 67.55 | 68.1 | 74.0 | 67.6 | 3.12 | 0.99 ± 0.02 |
| 0.41 | 296.9 | 124.6 | 68.3 | 69.18 | 70.0 | 76.9 | 69.3 | 4.27 | 1.46 ± 0.02 |
| 0.64 | 275.6 | 113.4 | 68.9 | 69.76 | 70.4 | 77.4 | 69.9 | 3.43 | 1.14 ± 0.02 |
| 0.87 | 259.3 | 108.3 | 69.1 | 70.2 | 70.6 | 77.9 | 70.3 | 3.57 | 1.16 ± 0.02 |
| 1.1 | 247.36 | 102.14 | 69.7 | 70.55 | 71.2 | 78.2 | 70.7 | 3.7 | 1.17 ± 0.02 |

**TABLE II.** Pseudogap parameters of $Y_{0.77}Pr_{0.23}Ba_2Cu_3O_{7-\delta}$ single crystal at different pressure.

| $P$, GPa | $T^*$, K | $C_{3D}$ | $A_4$ | $\alpha^*$ | $\varepsilon^*_{c0}$ | $T_{pair}$, K | $D^*$ | $\Delta^*(T_G)$, K | $\Delta^*(T_{pair})$, K |
|---|---|---|---|---|---|---|---|---|---|
| 0 | 163 | 0.7 | 16 | 2.5 | 0.4 | 105.1 | 5 | 183.0 | 151.9 |
| 0.41 | 166.0 | 1.45 | 28 | 2.7 | 0.37 | 107.8 | 5 | 166.3 | 159.6 |
| 0.64 | 166.3 | 1.075 | 23 | 2.8 | 0.36 | 107.4 | 5 | 170.0 | 152.0 |
| 0.87 | 166.0 | 1.1 | 25 | 2.8 | 0.36 | 106.7 | 5 | 179.0 | 151.9 |
| 1.1 | 164.3 | 1.2 | 26 | 2.9 | 0.34 | 106.1 | 5 | 174.2 | 148.0 |

Concurrently, the pressure-induced increase in the density of charge carriers $n_f$ in the $CuO_2$ planes should lead to an increase in $T_c$,[62–65] which is also observed in the experiment. As can be seen from Table I, $T_c$ increases from 66.4 K ($P = 0$) to 69.7 K ($P = 1.1$ GPa) (red dots in left inset in Fig. 1). Thus, in the single crystal YPrBCO under study, we have $dT_c/dP = +3.0\,KGPa^{-1}$. This is ~1.7 times less than in SD YBCO single crystals, where $dT_c/dP = +5.1\,KGPa^{-1}$,[51] but ~4.1 times higher than in OD of YBCO single crystals, where the increment rate of the critical temperature $dT_c/dP = +0.73\,KGPa^{-1}$.[52] It is also ~1.65 times higher than in the YPrBCO single crystal with $x = 0.05$, where the increment rate of the critical temperature $dT_c/dP = +1.82\,KGPa^{-1}$.[53] Thus, the localization of charge carriers due to the presence of PrBCO in this case has little effect on $T_c$. This result once again confirms that the mechanisms of the effect of hydrostatic pressure on the critical temperature and the resistivity of both YPrBCO and YBCO single crystals are most likely different.[21,50] The temperature of the resistive transition to the SC state, $T_c$ $(R = 0)$, was determined by extrapolation of the linear part of the SC transition to the intersection with the temperature axis,[20,40,41] as shown in the right inset to Fig. 1. The width of resistive transitions $\delta T_c = T_c(0.9\rho_n) - T_c(0.1\rho_n)$, where $\rho_n$ is the resistivity of the sample directly above the transition,[51] at $P = 0 \approx 1.2$ K, that is, it is quite small, suggesting good sample structure and absence of additional phases with different $T_c$. Surprisingly, at $P = 1.1$ GPa, still $\delta T_c \approx 1.2$ K. Thus, in this case, pressure has virtually no effect on $\delta T_c$, i.e., $d\delta T_c/dP = 0$. Usually, the pressure increases the width of resistive transitions. This effect is most pronounced in SD YBCO single crystals ($T_c(P = 0) = 49.2$ K), where $d\delta T_c/dP \approx 0.65\,KGPa^{-1}$[51] and especially in HoBCO single crystals ($T_c(P = 0) = 61.3$ K) with an effective magnetic moment of Ho $m_{eff} = 9.7\mu_B$ and containing prolonged defects in the form of TBs. In the latter case, $d\delta T_c/dP \approx 3.5\,KGPa^{-1}$.[50] Interestingly, in the case of the YPrBaCuO single crystal with $x = 0.05$, the lowest value of $d\delta T_c/dP \approx 0.35\,KGPa^{-1}$ is observed. Thus, our result indicates a tendency: the higher the Pr content, the less the influence of pressure on the width of the resistive transition. Resistive parameters of the samples under study at various pressures are given in Table I.

However, no peculiarities are observed in the $\rho(P)$ and $T_c(P)$ dependences (Fig. 1), and, strictly speaking, the reason for the linear decrease in $\rho(P)$ in the YPrBCO single crystals remained unclear. There is also no clear answer to the question of how the fluctuating Cooper pairs and PG behave under the influence of pressure. Thus, to obtain new information, we studied the pressure dependences of FLC and PG in the YPrBCO single crystal, as described in the following sections.

### 3.2. Fluctuation conductivity

One of the main properties of HTSC is the linear temperature dependence of the resistivity $\rho(T)$ in the normal state, that is, at $T > T^*$. However, as noted above, at $T \leq T^*$, $\rho(T)$ deviates downward from linearity, which leads to the occurrence of excess conductivity:

$$\sigma'(T) = \sigma(T) - \sigma_N(T) = 1/\rho(T) - 1/\rho_N(T), \tag{1}$$

where $\rho_N(T) = \alpha T + \rho_0$ is the resistivity of the sample in the normal state, extrapolated to the region of low temperatures (red lines in Fig. 1). As before, $\alpha = d\rho/dT$ determines the slope of the linear dependence $\rho_N(T)$, and $\rho_0$ is the residual resistance.[60,66,67] It should also be remembered that at $T \leq T*$, not only does $\rho(T)$ begin to deviate from linearity, but also the density of states at the Fermi level begins to gradually decrease, indicating the opening of the PG.[3,6–8,21] Furthermore, it is assumed that the Fermi surface also changes,[1–3,68] most likely due to the formation of LPs below $T^*$,[1,3,10–12] as mentioned above. Clearly, pressure must somehow influence these processes.

Numerous studies[3,21,31,66] (and references therein) have convincingly shown that near $T_c$ $\sigma'(T)$ is always described by the Aslamazov–Larkin (AL) equation for three-dimensional (3D) fluctuation conductivity:[69]

$$\sigma'_{AL3D} = C_{3D}\frac{e^2}{32\hbar\xi_c(0)}\varepsilon^{-1/2}, \tag{2}$$

where $C_{3D}$ is the scaling factor.[37] This is due to the fact that near $T_c$ the coherence length along the $c$-axis $\xi_c(T) > d \sim 11.69$ Å for YPrBSO,[70] where $d$ is the unit cell size along the $c$-axis. As a result, the LPs penetrate into the entire volume of the sample. Moreover, there is a certain range of superconducting fluctuations $\Delta T_{fl} \sim 15$ K above $T_c$, where LPs in cuprates behave like fluctuating Cooper pairs[3,20,21,60] and references therein), but without long-range order (the so-called "short-range phase correlations").[10–12,33,34,71,72] This is a rather specific HTSC state, which may occur due to the short coherence length $\xi(T)$ in combination with the quasi-layered structure of cuprates.[3,10,71]

To use Eq. (2), it is necessary to determine the reduced temperature

$$\varepsilon = \left(T - T_c^{mf}\right)/T_c^{mf}, \tag{3}$$

which is included in all equations. In turn, it depends on the critical temperature in the mean-field approximation, $T_c^{mf}$, which turns out to be an important parameter in both FLC and PG analysis. To find $T_c^{mf}$ we use an approach proposed in Ref. 37. From Eq. (2), it is clear that $\sigma'(T) \sim \varepsilon^{-1/2}$. Consequently, $\sigma'(T)^{-2} \sim \varepsilon \sim T - T_c^{mf}$. Finely, $\sigma'(T)^{-2} = 0$ when $T = T_c^{mf}$. Figure 2 shows the method for determining $T_c^{mf}$ for S1 ($P = 0$, blue dots) and S5 ($P = 1.1$ GPa, magenta dots). The straight line corresponds to the range of 3D–AL fluctuations. Its intersection with the $X$-axis just determines $T_c^{mf}$ = 67.55 K ($P = 0$) and = 70.55 K ($P = 1.1$ GPa). Also shown are $T_c$, the Ginzburg temperature $T_G$, down to which the mean-field theory works, and the 3D–2D crossover temperature $T_0$. Above $T_0$, the data deviate to the right of the line, indicating the presence of two-dimensional (2D) SC fluctuations of the Maki–Thompson (MT) type in the sample.

Having determined $T_c^{mf}$ and hence ε, we plot ln σ′ as a function of ln ε (Fig. 3). As expected, near $T_c$ the experimental data are well described by the 3D–AL fluctuation theory (Eq. (2), red solid lines with a slope of $\lambda = -1/2$, labeled 3D–AL in the figure). However, since above $T_c$ the coherence length is defined as $\xi(T) = \xi(0)\varepsilon^{-1/2}$,[31,73] it decreases rapidly with increasing $T$. Ultimately, above the crossover temperature $T_0$ (Table I), $\xi_c(T) < d$, and the 3D–AL fluctuation state is lost. Nevertheless, $\xi_c(T) > d_{01}$ is still the case, where $d_{01}$ (≈ 3.4 Å for YPrBCO[70] is the distance between the conducting planes of $CuO_2$ in HTSC. As a result, up to $T_{01}$ (Table I), $\xi_c(T)$ still connects the planes by Josephson interaction, forming a two-dimensional (2D) fluctuation state. In well-structured samples,[31,74] the 2D fluctuation state is described by the 2D Maki–Thompson (2D–MT) theory, modified by Hikami and Larkin (HL) for HTSCs:[75]

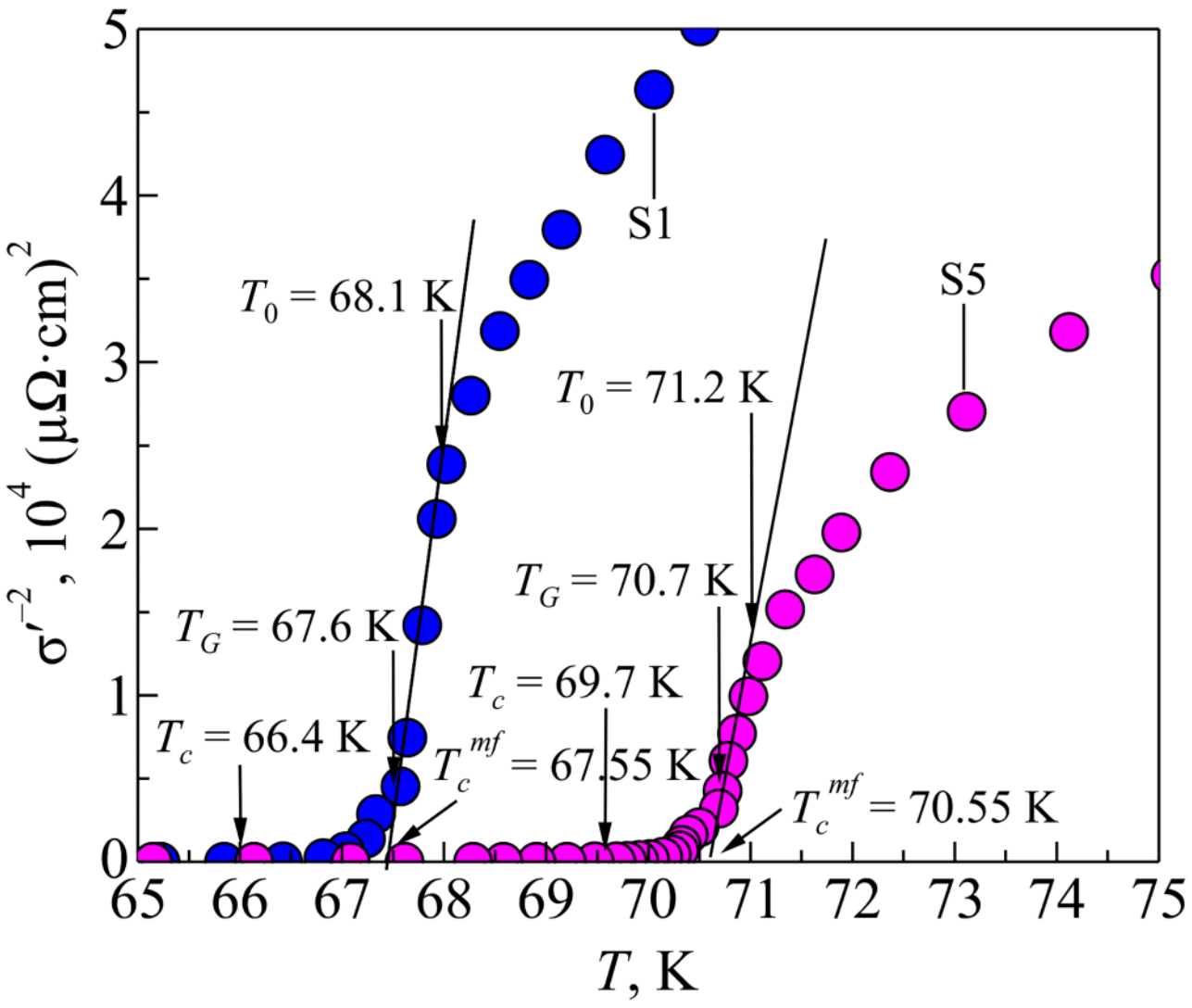

**FIG. 2.** $\sigma'^{-2}$ vs $T$ for the $Y_{0.77}Pr_{0.23}Ba_2Cu_3O_{7-\delta}$ single crystal at $P = 0$ (S1, blue dots) and at $P = 1.1$ GPa (S5, magenta dots). Arrows point to the characteristic temperatures: $T_c$, the mean-field transition temperature $T_c^{mf}$, the Ginzburg temperature $T_G$, and the 3D–2D crossover temperature $T_0$. Straight lines correspond to the range of 3D–AL fluctuations.

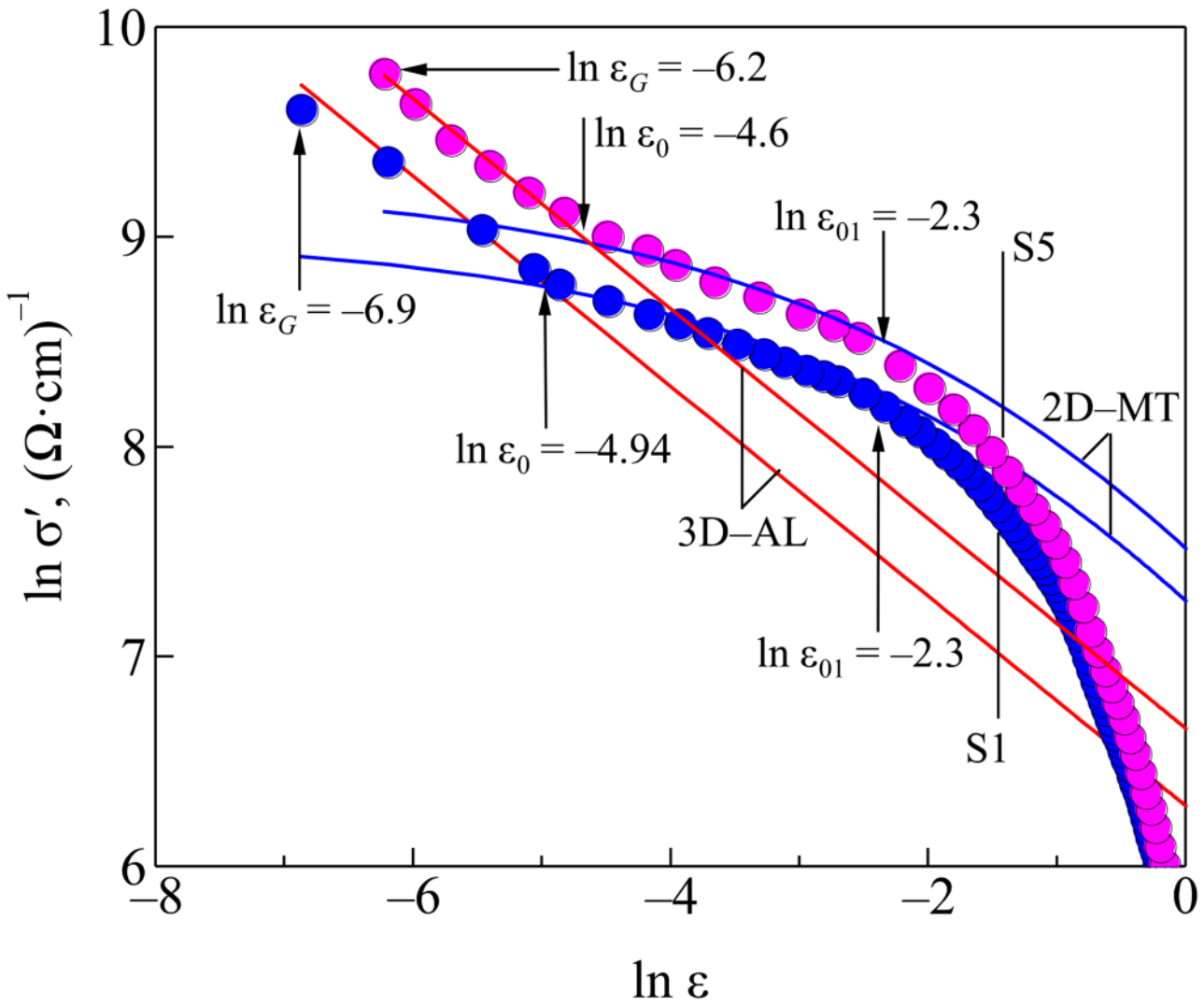

**FIG. 3.** ln σ′ vs ln ε for $Y_{0.77}Pr_{0.23}Ba_2Cu_3O_{7-\delta}$ single crystal at $P = 0$ (blue dots) and $P = 1.1$ GPa (magenta dots) in comparison with fluctuation theories: 3D–AL (red lines), 2D–MT (blue curves). All characteristic temperatures are marked with arrows: ln $\varepsilon_G$ refers to the Ginzburg temperature $T_G$, ln $\varepsilon_0$ to the 3D–2D crossover temperature $T_0$, and ln $\varepsilon_{01}$ to $T_{01}$, which limits the region of the SC fluctuations from above.

$$\sigma'_{MT} = C_{2D}\frac{e^2}{8d\hbar}\frac{1}{1-\alpha/\delta}\ln\left((\delta/\alpha)\frac{1+\alpha+\sqrt{1+2\alpha}}{1+\delta+\sqrt{1+2\delta}}\right)\varepsilon^{-1}, \tag{4}$$

where $\alpha = 2[\xi_c(0)/d]^2\varepsilon^{-1}$ is the coupling parameter,

$$\delta = \beta\frac{16}{\pi h}\left[\frac{\xi_c(0)}{d}\right]^2 k_B T\tau_\phi. \tag{5}$$

is the pair-breaking parameter. In accordance with the HL theory at the 3D–2D crossover temperature $T_0 \sim \varepsilon_0$, $\delta \cong \alpha$, which allows us to obtain an equation for the lifetime of the FCPs $\tau_\phi\beta T = \pi h/8k_B\varepsilon_0 = A/\varepsilon_0$, where $A = 2.998\cdot 10^{-12}$ s·K. The factor $\beta = 1.203(l/\xi_{ab})$, where $l$ is the mean free path, and $\xi_{ab}$ is the coherence length along $CuO_2$ planes, corresponds to the case of the clean limit ($l > \xi$) being typical for HTSCs.[31,76] In accordance with our consideration (see Fig. 3), at $T = T_0$ (ln $\varepsilon_0$ in the figures), which limits the range of 3D fluctuations from above, a clear crossover to the region of 2D fluctuations is observed. In this case above $T_0$ σ′ (ε) for both samples is well described by 2D–MT fluctuation contribution calculated using [Eq. (4)] (Fig. 3, solid blue curves), which is typical for well-structured HTSCs.[31,74] This fact suggests that the number of defects created by PrBCO is not that great.

It is clear that at $T = T_0$, $\xi_c(T_0) = \xi_c(0)\varepsilon_0^{-1/2} = d$ and $\xi_c(0)$ can be easily determined by the simple formula[5,60]

$$\xi_c(0) = d\sqrt{\varepsilon_0}. \tag{6}$$

The $\xi_c(0)$ values found using Eq. (6) for all five pressure values (five samples) are given in Table I. It can be seen that $\xi_c(0)$ increases slightly with increasing pressure, which is typical for YBCO[44,52] and YPrBCO[53] with high $T_c$. The only exception is S2 ($P = 0.41$ GPa), which shows a sharp increase in $\xi_c(0)$ (=1.45 Å) and $d_{01}$ (=4.27 Å). Concurrently, $C_{3D}$ also increases to 1.45 (Table II), whereas in well-structured YBCO, $C_{3D}$ should be around 1.[60,74] These results indicate that at a relatively small $P = 0.41$ GPa, PrBCO impurities create numerous defects in the crystal.

The 2D state is maintained until, at $T_{01}$, above which $\xi_c(T) < d_{01}$, the experimental data completely deviate from the theory[31,73] (and references therein). This occurs because in this case, any correlation interaction between the layers is lost, and the LPs are confined in the conducting planes of $CuO_2$. Thus, $T_{01}$ is the temperature that limits the range of SC fluctuation $\Delta T_{fl} = T_{01} - T_G$. So, it turns out that fluctuation conductivity is only part of the complete $\sigma'(T)$ given by Eq. (1) and is realized in a relatively narrow (~15 K) range of SC fluctuations above $T_c$. Accordingly, at $T = T_{01}\xi_c(T_{01}) = \xi_c(0)\varepsilon_{01}^{-1/2} = d_{01}$. Since $\xi_c(0)$ has already been determined by $\varepsilon_0$ [Eq. (6)], a simple algebra yields $d_{01} = d\sqrt{\varepsilon_0}/\sqrt{\varepsilon_{01}}$ (Table II). The values of $d_{01}$ calculated using this simple formula are given in Table I. With increasing pressure $P$, the value of $d_{01}$ smoothly increases from 3.12 Å ($P = 0$) to 3.7 Å at $P = 1.1$ GPa, which is comparable with the Cu(2)–Cu(2) distance = 3.44 Å between Cu atoms in adjacent $CuO_2$ planes in YPrBCO.[70] The very small value of $d_{01} = 3.12$ Å at $P = 0$ indicates the influence of magnetic impurities of PrBCO on the structure of the sample. The observed decrease in $d_{01}$ at $P = 0$ is possibly due to the increase in the Cu(2)–Cu(1) distance between the $CuO_2$ planes and Cu–O chains[70] under the influence of Pr magnetism.[45] However, as before, a feature of $d_{01} = 4.28$ Å at $P = 0.41$ GPa is observed, which indicates a strong influence of magnetic impurities of PrBCO at this pressure. The observed increase in $d_{01}$ at moderate pressure indicates that such pressure somehow increases the influence of disordered magnetic moments of PrBCO. This conclusion is supported by the fact that $C_{3D}$ at $P = 0.41$ GPa exhibits the largest value of 1.45, as mentioned above. A further moderate increase in $d_{01}$ (Table I) suggests a change in the mechanism of interaction between charge carriers in HTSC and magnetic impurities with increasing pressure, which is confirmed by a decrease in the $C_{3D}$ factor to almost unity. It should be emphasized that all the above-considered features of the behavior of FLCs in HTSCs, which look unexpected from the point of view of traditional superconductivity, are possible only due to the very short $\xi_c(T)$, which turns out to be comparable to the characteristic dimensions of the crystal structure d and $d_{01}$ along the $c$-axis. To obtain further information, a PG study was conducted at different $P$ as described in the next section.

### 3.3. Temperature dependence of pseudogap

Let us recall that the opening of the PG occurs at $T \leq T^*$ simultaneously with the deviation of the resistivity downwards from linearity and the appearance of excess conductivity.[3,60] Thus, it is obvious that $\sigma'(T)$ must contain information about PG. To obtain this information, an equation was proposed that describes $\sigma'(T)$ over the entire temperature range from $T^*$ to $T_G$ and includes PG explicitly:[60]

$$\sigma'(T) = A_4 \frac{e^2\left(1 - \dfrac{T}{T*}\right)\exp\left(-\dfrac{\Delta *}{T}\right)}{16\hbar\xi_c(0)\sqrt{2\varepsilon *_{c0}\sinh\left(2\dfrac{\varepsilon}{\varepsilon *_{c0}}\right)}}, \tag{7}$$

where $A_4$ is a numerical factor, the same as the $C$-factor in the FLC theory [see Eqs. (2) and (4)], and $\Delta^*$ is the value of a PG parameter at $T = T_G$.[3,60] Solving Eq. (7) with respect to $\Delta^*$, we obtain an equation for the PG:[60]

$$\Delta * (T) = T\ln\left[A_4\left(1 - \frac{T}{T*}\right)\frac{1}{\sigma'(\varepsilon)}\frac{e^2}{16\hbar\xi_c(0)} \times \left[\frac{1}{\sqrt{2\varepsilon *_{c0}\sinh(2\varepsilon/\varepsilon *_{c0})}}\right]\right]. \tag{8}$$

An important point is the fact that all parameters included in these equations can be determined from experiment. The values of $T^*$, $\varepsilon$ [Eq. (3)], and $\xi_c(0)$ [Eq. (6)] have already been determined from the resistivity and the FLC analysis and are summarized in Tables I and II. Other necessary parameters, such as $\Delta^*(T_G)$, the theoretical parameter $\varepsilon^*_{c0}$,[77] and the coefficient $A_4$, can now be directly derived from the PG analysis.[3,60]

It has been found that in all HTSCs $\sigma'^{-1} \sim \exp(\varepsilon)$ in a certain temperature range $\ln\varepsilon_{c01} < \ln\varepsilon < \ln\varepsilon_{c02}$ above $T_{01}$ (arrows in Fig. 4).[20,31,50,74,77] Simple algebra yields $\ln(\sigma'^{-1}) \sim \varepsilon$. Consequently, in the temperature range $\varepsilon_{c01} < \varepsilon < \varepsilon_{c02}$, $\ln(\sigma'^{-1})$ is a linear function of $\varepsilon$ with a slope $\alpha^* = 2.5$, which defines the parameter $\varepsilon^*_{c0} = 1/\alpha^* = 0.4$ at $P = 0$ (insets to Fig. 4). This approach allows the determination of reliable values of $\varepsilon^*_{c0}$ for all pressures (see Table II), which significantly affect the curvature of the theoretical curves shown in (Figs. 4 and 5) at high temperatures[31,60,74] and slightly increases with $P$. Next, to find $\Delta^*(T_G)$, we plot the experimental excess conductivity in coordinates $\ln\sigma'$ versus $1/T$[31,50,60] and approximate them by the $\ln\sigma'(1/T)$ calculated from the Eq. (7). Figure 5 shows two such plots for $P = 0$ (S1, blue dots) and $P = 1.1$ GPa (S5, magenta dots). The curves for S2 ($P = 0.41$ GPa), S3 ($P = 0.64$ GPa), and S4 ($P = 0.87$ GPa) look similar but are not shown to avoid cluttering the graph. With such a construction, the shape of the theoretical curves turns out to be very sensitive to the value of $\Delta^*(T_G)$.[20,31,60] It should be emphasized that in cuprates $\Delta^*(T_G) = \Delta(0)$, which is an SC energy gap at $T = 0$.[17,18] Taking this fact into account, we can determine the Bardeen–Cooper–Schrieffer (BCS) ratio $D^* = 2\Delta(0)/k_BT_c = 2\Delta^*(T_G)/k_BT_c$, and hence the value $\Delta^*(T_G)$ (Table II) at all pressures. At different $P$, the best approximation is achieved at the value of $D^* = 5.0$, which is a typical value for YBCO,[78] suggesting the strong coupling limit for our samples despite the strong influence of defects.

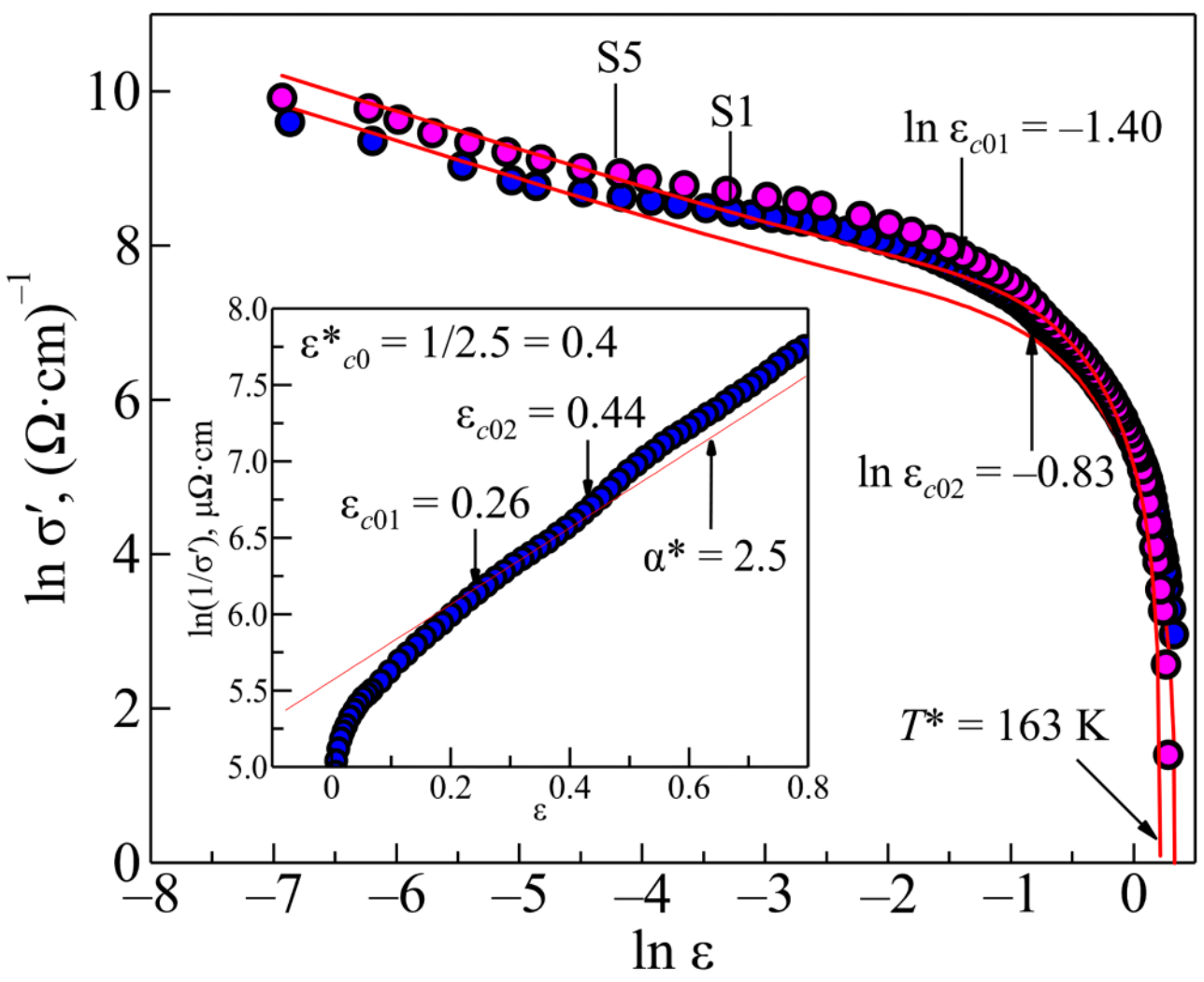

**FIG. 4.** ln $\sigma'$ vs ln $\varepsilon$ for $Y_{0.77}Pr_{0.23}Ba_2Cu_3O_{7-\delta}$ single crystal at $P = 0$ (S1, blue dots) and $P = 1.1$ GPa (S5, magenta dots), plotted over the entire temperature range from $T^*$ down to $T_c^{mf}$ and compared with our theoretical model [Eq. (7)] (red curves). The arrows designate the ranges of the dependences $\sigma'^{-1} \sim \exp(\varepsilon)$. Inset ln $\sigma'^{-1}$ as a function of $\varepsilon$. The solid line indicates the linear part of the curve between $\varepsilon_{c01}$ and $\varepsilon_{c02}$. Its slope $\alpha^*$ determines the parameter $\varepsilon^*_{c0} = 1/\alpha^*$.

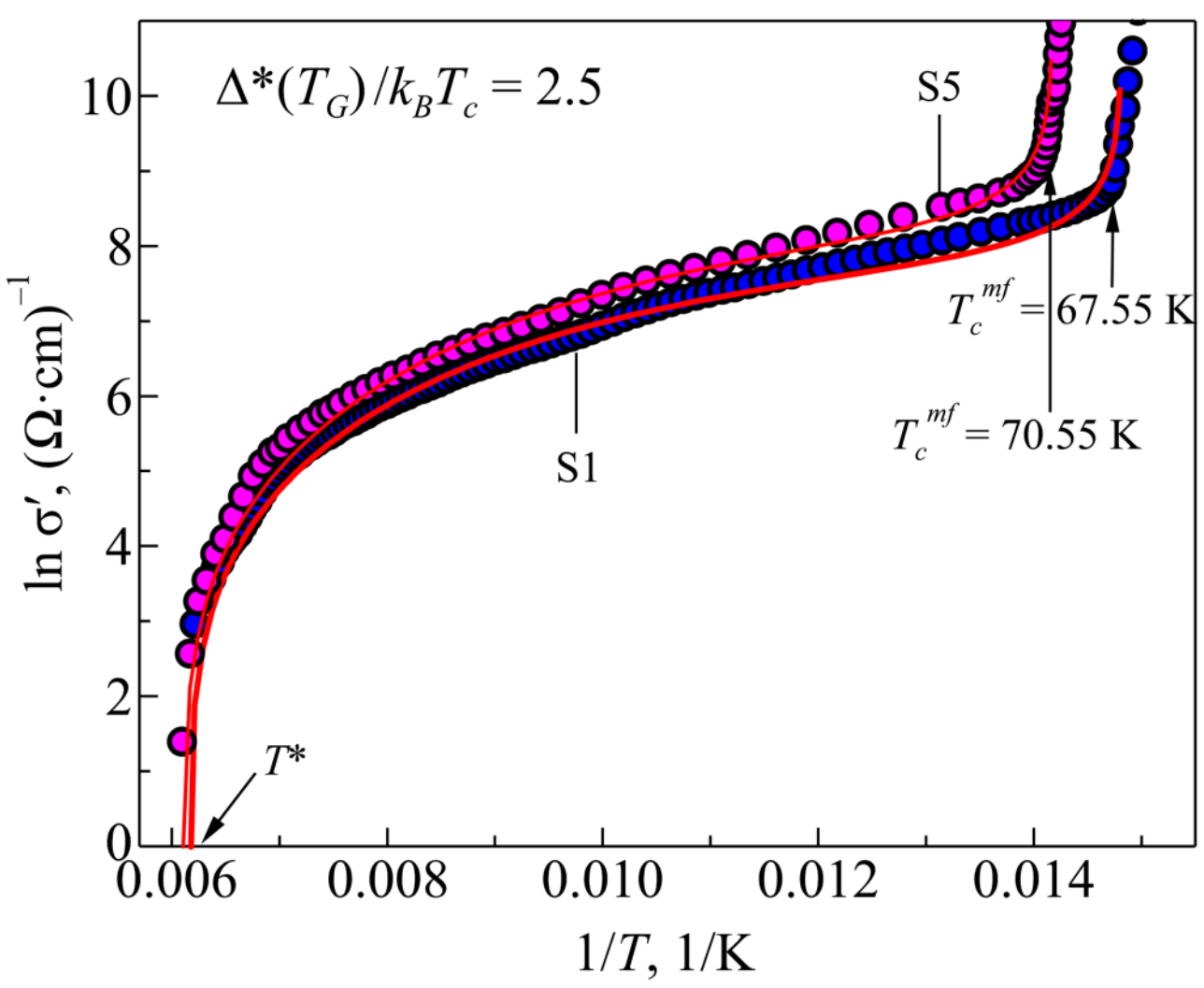

**FIG. 5.** ln $\sigma'$ vs $1/T$ for $Y_{0.77}Pr_{0.23}Ba_2Cu_3O_{7-\delta}$ single crystal at $P = 0$ (S1, blue dots), and $P = 1.1$ GPa (S5, magenta dots) plotted in the whole temperature range from $T^*$ down to $T_c^{mf}$ marked by arrows. The red curves are fits to the data with Eq. (7). The best fits are obtained when Eq. (7) is calculated with $D* = 2\Delta*(T_G)/k_BT_c = 5.0$ for all pressures.

And finally, to find the scale factor $A_4$, we plot experimental data of $\sigma'(T)$ for all pressures over the entire temperature range from $T^*$ to $T_G$ (Fig. 4) and approximate them with Eq. (7), calculated taking into account the already found parameters (the red curves in the figure).[3,31,50,60] As before, we show results only for $P = 0$ (blue dots) and $P = 1.1$ GPa (magenta dots). Now the only fitting parameter is $A_4$. By choosing $A_4$, we approximate the experiment with theoretical curves in the region of 3D−AL fluctuations, where ln $\sigma'$ is a linear function of ln $\varepsilon$ with slope $\lambda = -1/2$.[3,20,31,50,60,74] (Fig. 4, red curves). As can be seen in the figure, very good agreement between the data and our theoretical model is observed. The same results were obtained for all other samples. Interestingly, $A_4$ increases significantly from $A_4 = 16$ ($P = 0$) to 28 ($P = 0.41$ GPa), likely due to the increase in defects under applied pressure. However, at $P = 0.64$ GPa, $A_4$ unexpectedly decreases to 23, with $C_{3D}$ demonstrating the closest value to unity, 1.07 (Table II). This result suggests that such pressure somehow neutralizes the influence of defects created by magnetic inclusions of the PrBCO. With further increase in $P$, $A_4$ slightly increases to 25 with a simultaneous increase in $C_{3D}$ to 1.2 ($P = 1.1$ GPa), which can be regarded as an increase in the influence of defects. However, before drawing conclusions, the following result should be noted. In Fig. 4, it is clearly seen that at $P = 0$ (blue dots), the experiment significantly deviates upward from the theoretical curve in the region of 2D fluctuations. A sharp increase in 2D fluctuations was observed for all cuprates containing magnetic impurities[31] and especially for FeAs-based superconductors.[79] The deviation from the theory is explained by the fact that our theoretical model does not take into account the influence of magnetism.[60] At the same time, the deviation of the experimental dependence of ln $\sigma'$ vs ln $\varepsilon$ from the theory at $P = 1.1$ GPa (Fig. 4, magenta dots) is noticeably smaller, possibly due to a reduced influence of magnetic impurities. To clarify this unexpected result, the PG study appears to be very useful.

The temperature dependence of PG, $\Delta^*(T)$, at $P = 0$, calculated using Eq. (8) with the following set of parameters: $T^* = 163$ K, $T_c^{mf} = 67.5$ K, $\xi_c(0) = 0.99$ Å, $\varepsilon^*_{c0} = 0.4$, $A_4 = 16$, and $D^* = 5$, is shown in Fig. 6(a) (blue dots). As expected, $\Delta^*(T)$ shows features typical of YPrBCO with relatively high oxygen content.[32] Below $T^*$, $\Delta^*(T)$ rapidly increases, showing a relatively narrow maximum at $T_{pair} = 105.1$ K, followed by a pronounce minimum at $T_{01} = 74.0$ K and then increases sharply to ∼183 K at $T_G = 67.6$ K. It is interesting that in this case the minimum at $T_G$ is absent, most likely due to the influence of the magnetic impurities PrBCO, but $T_G$ can be determined from Fig. 2. Additional feature is a small magnetic maximum at $T_m \approx 150$ K. In Refs. 31 and 32, it has been convincingly demonstrated that with increasing Pr content, the maximum rapidly increases, and the entire $\Delta^*(T)$ dependence assumes a shape similar to that observed in magnetic superconductors.

In this study, the shape of $\Delta^*(T)$ at different $P$ is almost the same, but with increasing $P$, the magnetic maximum is noticeably suppressed and almost completely disappears at $P = 0.87$ GPa. This result shows that pressure somehow reduces the influence of magnetic moments of PrBCO at $P = 0.87$ GPa, resulting in the largest $\Delta^*(T_G)/k_B = 179.0$ K and $C_{3D}$ close to unity (Table II). However, oddly enough, at $P=1.1$ GPa the magnetic maximum is restored and even becomes more pronounced than at $P=0$ (Fig. 6(b), magenta dots). Indeed, the length of the linear part of this maximum along the $Y$-axis is $\Delta_Y \approx 10.1$ K ($P=0$) and $\approx$22.7 K at $P=1.1$ GPa. Moreover, as can be seen from the figure, the slope of this linear part $\alpha=4.3$ does not depend on the pressure. Nevertheless, at $P=1.1$ GPa, $\Delta^*(T_G)/k_B$ unexpectedly decreases to 174.2 K, and $\Delta^*(T_{pair})/k_B$ demonstrates the lowest value of 148.0 K (Table II). This indicates changes in the mechanisms of interaction of the charge carrier subsystem with magnetic impurities in the sample with increasing $P$ above ~1 GPa.

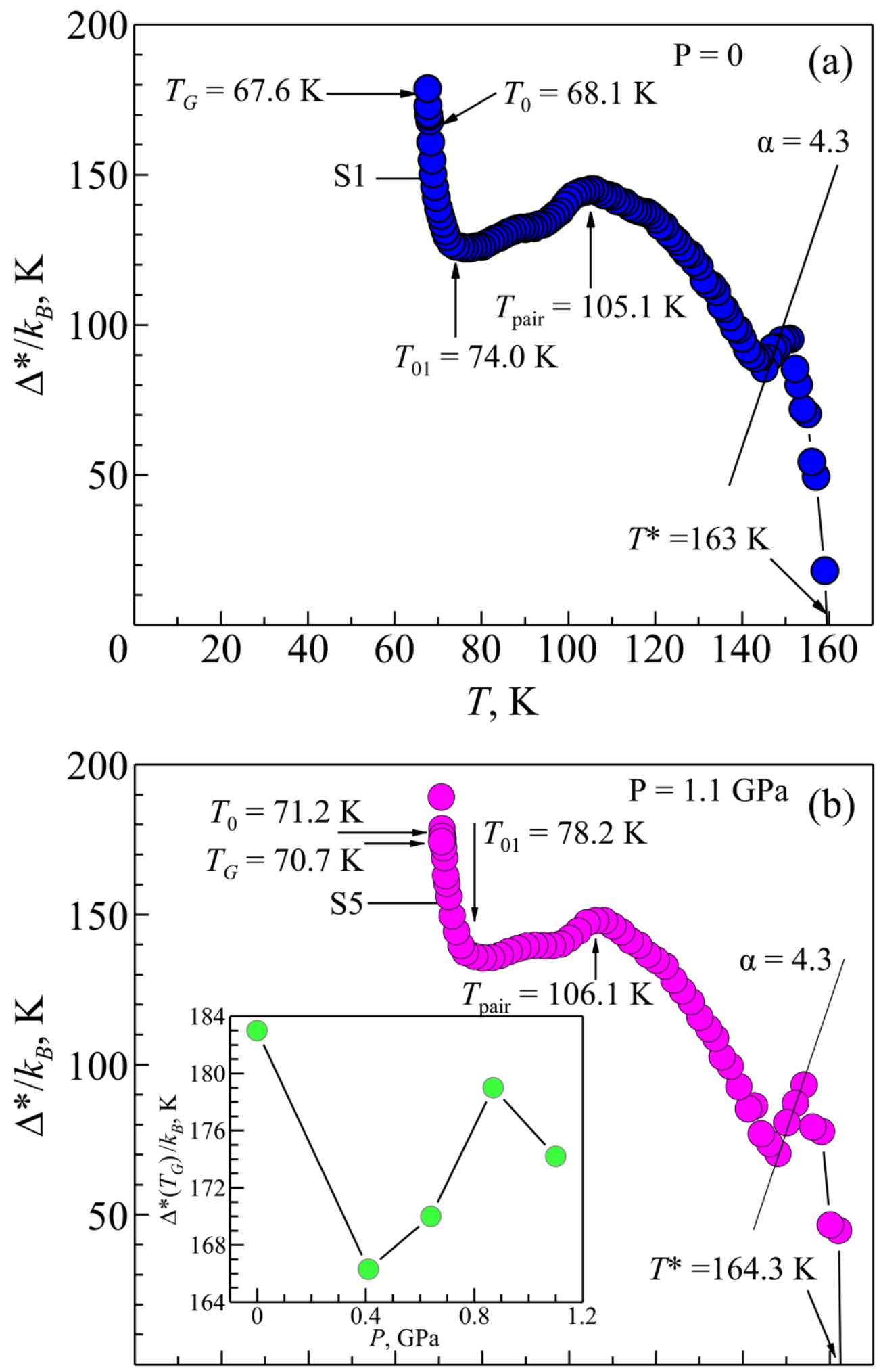

**FIG. 6.** Temperature dependences of pseudogap $\Delta^*/k_B$ for $Y_{0.77}Pr_{0.23}Ba_2Cu_3O_{7-\delta}$ single crystal (a) at $P=0$ (S1, blue dots) and (b) $P=1.1$ GPa (S5, magenta dots) calculated using Eq. (8) with the parameters given in the text. The arrows indicate the corresponding characteristic temperatures. Inset shows dependence $\Delta^*(T_G)$ vs $P$ in the range from $P=0$ to $P=1.1$ GPa. Solid curves are a guide for the eyes.

This conclusion is confirmed by the pressure dependence of $\Delta^*(T_G)/k_B$ (see inset to Fig. 6, green dots). This simple graph, together with our previous considerations, suggests the following scenario. Relatively low pressure promotes the formation of defects caused by PrBCO magnetic inclusions. As a result, features are observed in $\xi_c(0)$ and $d_{01}$ (Table I), accompanied by the highest value of $C_{3D}=1.45$ at $P=0.41$ GPa. At the same time, $\Delta^*(T_G)/k_B$ drops sharply to its lowest value of 166.3 K at this pressure, presumably due to the strong influence of defects (Table II). However, starting from 0.6 GPa, the mechanism of interaction between the electron subsystem of charge carriers and the magnetic defects of PrBCO changes, and pressure, as noted above, somewhat neutralizes the influence of magnetic impurities. As a result, $\Delta^*(T_G)/k_B$ recovers its value up to 179.0 K at $P=0.87$ GPa, $C_{3D}$ is close to unity, and the magnetic maximum disappears. However, starting from ~1 GPa, the mechanism apparently changes again, and $\Delta^*(T_G)/k_B$ suddenly decreases to 174.2 K (Table II). It appears that now the pressure, firstly, somehow orders the defects and restores the behavior of $\Delta^*(T)$ typical for most HTSCs near $T_c$[31,32] with a pronounced minimum at $T_{01}$, a maximum just below $T_0$ and a small minimum at $T_G$, which is completely suppressed by defects at $P=0$ (Fig. 7). Secondly, pressure apparently aligns the magnetic moments of PrBCO. As a result, a magnetic maximum reappears on $\Delta^*(T)$, more pronounced than at $P=0$. Moreover, the slope of the linear portion of this maximum, $\alpha=4.3$, does not change with pressure, as mentioned above (Fig. 6(b), magenta dots). In Fig. 8, taken from Ref. 32, we compared this slope with the same slope of the $Y_{0.57}Pr_{0.43}Ba_2Cu_3O_{7-\delta}$ single crystal and other magnetic HTSCs and showed that the slope is the same, which confirms the magnetic nature of the maximum observed in our experiments.

Finally, we compared found $\Delta^*(T)$ of samples S1 and S5 near $T_c$ with the temperature dependences of the local pair density in HTSCs $<n_\uparrow n_\downarrow>$ calculated in the Peters–Bauer theory within the framework of the three-dimensional attractive Hubbard model for different temperatures $T/W$, interactions $U/W$, and filling factor, where $W$ is the band width Ref. 12. This makes it possible to estimate the value $<n_\uparrow n_\downarrow>$ in both S1 and S5 sample at $T_G$, that is, to determine the density of local pairs in the studied single crystal depending on pressure (Fig. 9). Since the theoretical curves exhibit similar behavior with a maximum and minimum near $T_c$ as S5, we compare the measured values of normalized $\Delta^*/\Delta^*_{max}$ for S5 at $T_G$ with the minimum and at $T_0$ with the maximum of each theoretical curve calculated at different $U/W$ values, thereby achieving the best agreement between experiment and theory over the widest possible temperature range. It is important that the fitting coefficients found for S5 also be used for S1. The fitting results for the three $U/W$ values are shown in Fig. 9.

Sample S5 (magenta dots) shows excellent agreement with the theory at $U/W=0.6$ (small green dots) over a relatively large range of SC fluctuation above $T_0$. This fit corresponds to the value $<n_\uparrow n_\downarrow>(T_G) \approx 0.4$. This result supports the conclusion that pressure above 1.0 GPa somehow aligns the magnetic moments of PrBCO. On the other hand, due to the influence of disordered magnetic impurities of PrBCO, the behavior of S1 ($P=0$) turned out to be much more unusual. Using the same fitting coefficients as for S5, we were able to align the maximum of $\Delta^*/\Delta^*_{max}$ for S1 just below $T_0$ with the corresponding maximum on the theoretical curve at $T/W = 0.4$ (small orange dots). In this case, the experimental data in a wide range of temperatures above $T/T^* \approx 0.2$ run parallel to the theory at $T/W = 0.4$ and are very likely to be consistent with the theory at $T/W \approx 0.3$, which is not shown. However, below $T/T^* = 0.18$, the data begin to increase noticeably. At the maximum below $T_0$ the experimental data points show a downward trend, but then continue to increase up to $T_G$, which, as noted above, is defined as the last point on the linear dependence in Fig. 2. As already noted, in this case the expected minimum at $T_G$, characteristic of well-structured HTSCs,[31,32,74] is absent, which is most likely due to the influence of disordered magnetic impurities. Ultimately, there is uncertainty in the local charge carrier density in HTSC $< n_\uparrow n_\downarrow >$ in this case. The data above $T/T^* \approx 0.2$ corresponds to $< n_\uparrow n_\downarrow > \approx 0.32$. At the same time, the actually measured but distorted $T_G$ value corresponds to $< n_\uparrow n_\downarrow > \approx 0.38$. However, the most important fact is that pressure clearly increases $< n_\uparrow n_\downarrow >$ to 0.4 for S5. Thus, this analysis confirms the above conclusion that pressure increases the charge carrier density in HTSC.[45]

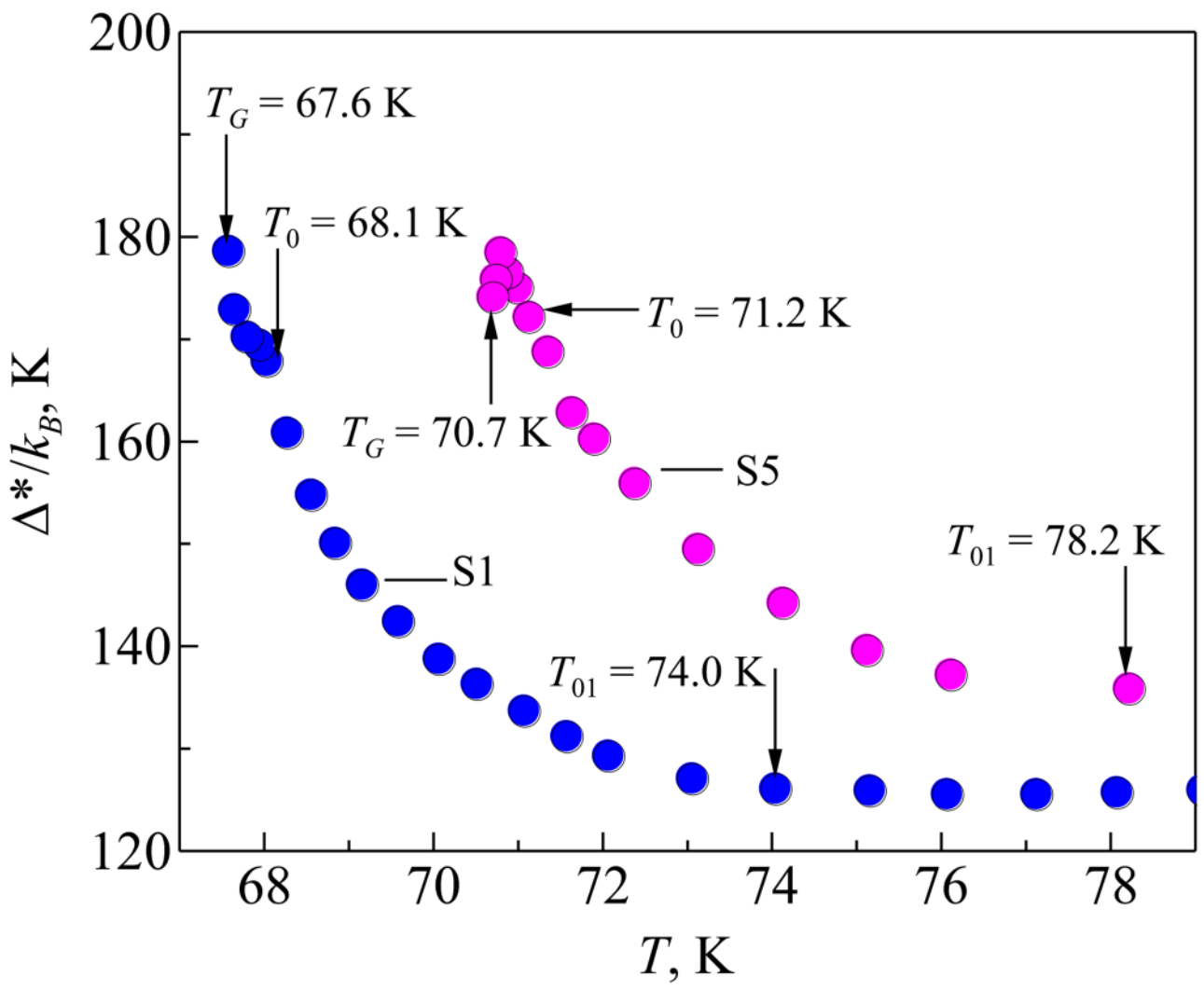

**FIG. 7.** Temperature dependences of pseudogap $\Delta^*/k_B$ for $Y_{0.77}Pr_{0.23}Ba_2Cu_3O_{7-\delta}$ single crystal at $P = 0$ (S1, blue dots) and $P = 1.1$ GPa (S5, magenta dots) calculated using Eq. (8) near $T_c$. At $P = 0$, the minimum at $T_G$ is completely destroyed by defects. The characteristic temperatures are designated by arrows.

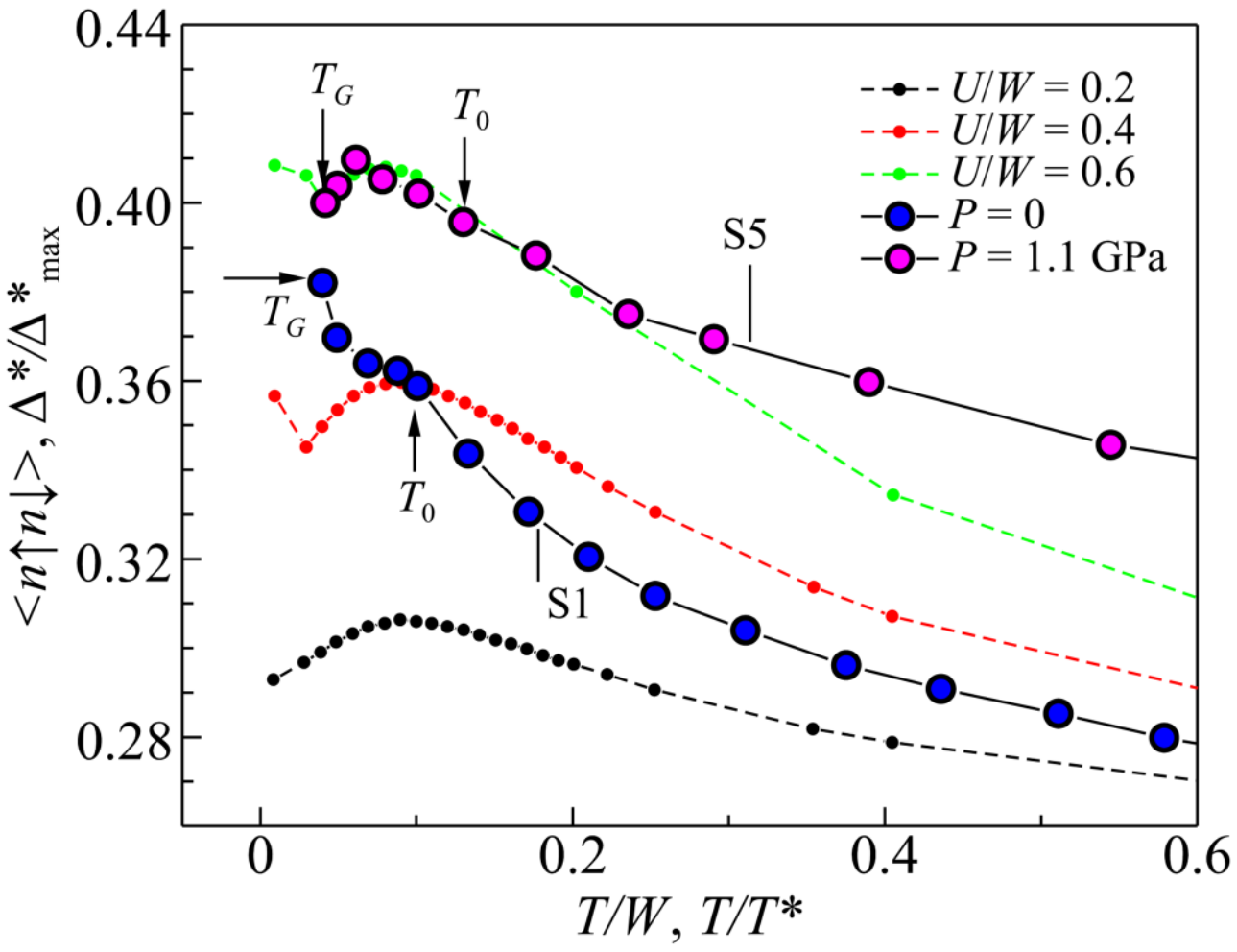

**FIG. 9.** The curves with small dots display the theoretical density of local pairs $< n\uparrow n\downarrow >$ as a function of $T/W$ for different values of the interaction energy: $U/W = 0.2$ (black dots), $U/W = 0.4$ (orange dots), and $U/W = 0.6$ (green dots). The big dots are experimental data for S1 ($P = 0$, blue dots) and S5 ($P = 1.1$ GPa, magenta points), after corresponding renormalizations of the dependences $\Delta^*(T)/\Delta^*_{max}$ in the interval $T_G < T < T_{01}$ as a function of reduced tempearture $T/T^*$ (see text). Solid curves between the dots are a guide for the eyes.

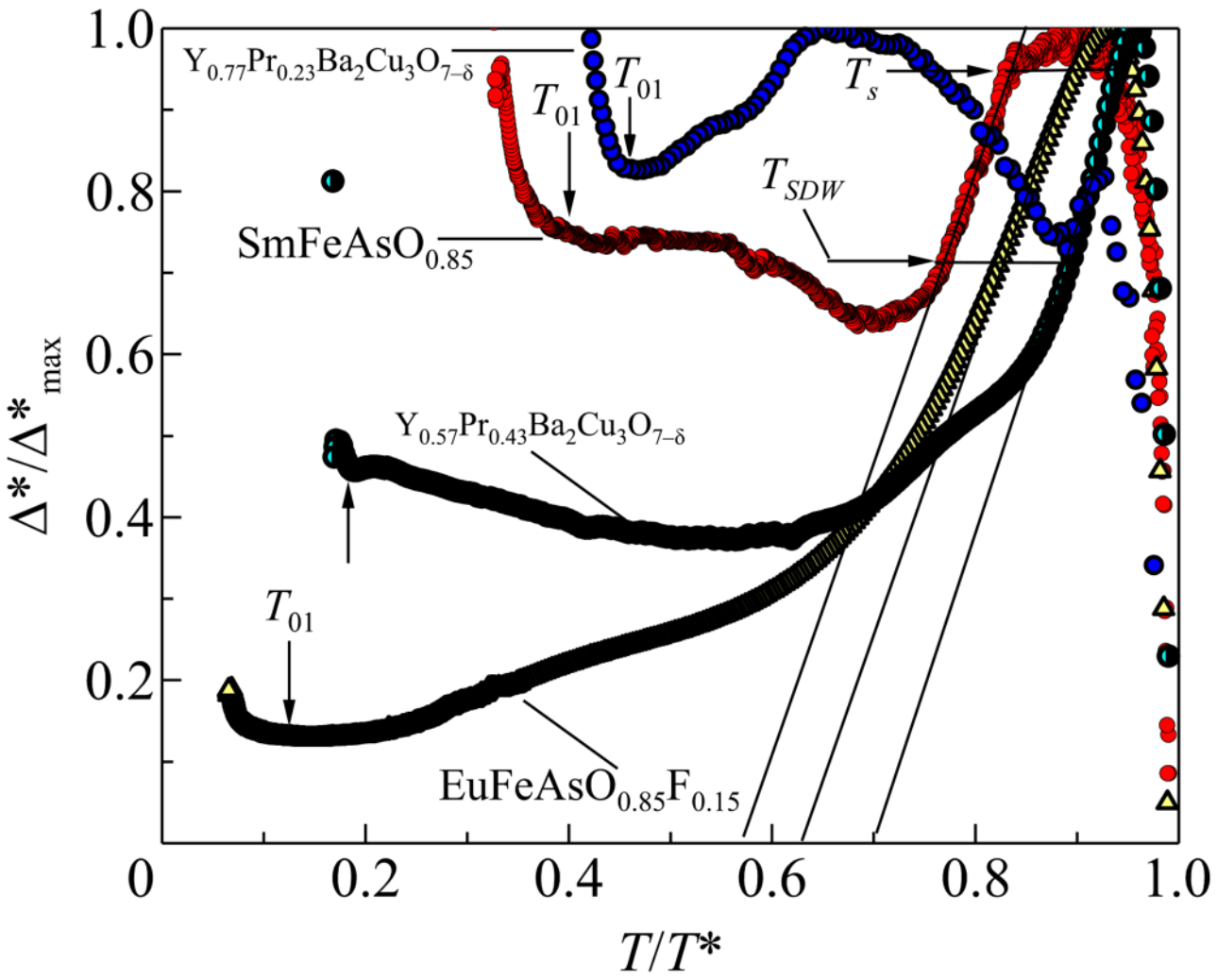

**FIG. 8.** $\Delta^*(T)/\Delta^*_{max}$ as a function of $T/T^*$ for the $Y_{0.77}Pr_{0.23}Ba_2Cu_3O_{7-\delta}$ single crystal (blue dots) compared with $Y_{0.57}Pr_{0.43}Ba_2Cu_3O_{7-\delta}$ single crystal and FeAS-based superconductors $SmFeAsO_{0.85}$ and $EuFeAsO_{1-x}F_x$ (Ref. 32). The slope of the linear section $\Delta^*(T)/\Delta^*_{max}$ below the magnetic maximum between temperatures $T_s$ and $T_{SDW}$ is practically the same for all samples.

## 4. CONCLUSION

In this paper, we studied the effect of hydrostatic pressure on resistivity $\rho(T)$, superconduting transition tempetature $T_c$, fluctuation conductivity (FLC), and pseudogap (PG) in $Y_{1-x}Pr_xBa_2Cu_3O_{7-\delta}$ ($x = 0.23$) single crystal. As expected, it was found that pressure significantly reduces resistance as $d\ \ln\ \rho(100\ \mathrm{K})/dP = -(29 \pm 0.2)\ \%\cdot\mathrm{GPa}^{-1}$, which is unexpectedly large compare with $d\ \ln\ \rho(100\ \mathrm{K})/dP = -(10.6 \pm 0.2)\ \%\cdot\mathrm{GPa}^{-1}$ obtained by us for YPrBCO single crystal with $x = 0.05$.[53] Thus, increasing the Pr content obviously enhances the effect of pressure on resistivity. This decrease is most likely due to an increase in the charge carrier density $n_f$ in the $CuO_2$ planes under pressure.[45] The same pressure-induced increase in $n_f$ leads to an observed increase in $T_c$ at a rate $dT_c/dP = +3.0$ KGPa$^{-1}$. This is ~1.65 times higher than in the YPrBCO single crystal with $x = 0.05$, where the increment rate of the critical temperature $dT_c/dP = +1.82$ KGPa$^{-1}$.[53] Thus, the localization of charge carriers due to the presence of PrBCO in this case has little effect on $T_c$. It was also shown that the width of resistive transitions $\delta T_c = T_c(0.9\rho_n) - T_c(0.1\rho_n)$, where $\rho_n$ is the resistivity of the sample directly above the transition, at $P = 0$ $\approx 1.2$ K, and practically does not change with pressure. Thus, in this case, pressure has virtually no effect on $\delta T_c$, i.e., $d\delta T_c/dP = 0$. Usually, the pressure increases the width of resistive transitions. Interestingly, among the cuprates that we studied, the lowest value of $d\delta T_c/dP \approx 0.35$ KGPa$^{-1}$ was observed for the YPrBaCuO single crystal with $x = 0.05$. Thus, our result indicates a tendency: the higher the Pr content, the less the influence of pressure on the width of the resistive transition.

However, resistance measurements provide no information about the charge carrier subsystem in the sample and its interaction with the magnetic impurities of PrBCO. To obtain this information, we analyzed the FLC and PG at various pressures. As a result, we were able to discover that the mechanism of interaction of charge carriers with magnetic impurities changes several times with increasing pressure. From the FLC analysis, we were able to measure the coherence length along the $c$-axis, $\xi_c(0)$, the $C_{3D}$ scaling factor, and estimate the distance between the $CuO_2$ conducting planes, $d_{01}$, of the $Y_{0.77}Pr_{0.23}Ba_2Cu_3O_{7-\delta}$ single crystal at different pressures. Comparing these parameters with the results of the PG analysis, we identified the features of $\xi_c(0)$ and $d_{01}$ with the largest $C_{3D} = 1.45$ at $P = 0.41$ GPa. At the same time, $\Delta^*(T_G)/k_B$ drops sharply to its lowest value of 166.3 K at this pressure, presumably due to the strong influence of defects. Thus, we can conclude that relatively low pressure promotes the formation of defects caused by PrBCO magnetic inclusions. However, starting from 0.6 GPa, $\Delta^*(T_G)/k_B$ recovers its value up to 179.0 K at $P = 0.87$ GPa, $C_{3D}$ becomes close to unity, and the magnetic maximum at $\Delta^*(T)$ disappears. Thus, a second conclusion can be drawn: the mechanism of interaction between the electron subsystem of charge carriers and the magnetic defects of PrBCO has changed such that pressure somewhat neutralizes the influence of magnetic impurities.

Finally, we may conclude, that, starting from ~1 GPa, the mechanism apparently changes again, and $\Delta^*(T_G)/k_B$ suddenly decreases to 174.2 K. It looks like, that the pressure at $P > 1.0$ GPa somehow orders the defects and restores the behavior of $\Delta^*(T)$ typical for most HTSCs near $T_c$ with a pronounced minimum at $T_{01}$, a maximum just below $T_0$ and a small minimum at $T_G$, which is completely suppressed by defects at $P = 0$ (Fig. 7). Secondly, pressure appears to equalize the magnetic moments of PrBCO. As a result, a magnetic maximum reappears on $\Delta^*(T)$, even more pronounced than at $P = 0$, with the same slope $\alpha = 4.3$, which does not change with pressure. Thus, analysing the temperature dependences of PG, $\Delta^*(T)$, of the $Y_{0.77}Pr_{0.23}Ba_2Cu_3O_{7-\delta}$ single crystal, we were able to obtain new and unexpected results. Really, we have shown that the mechanism of interaction between the subsystem of charge carriers and the magnetic defects of PrBCO in the crystal can change with the increase of pressure. Furthermore, the comparison with the Peters–Bauer theory showed that the local pair density in HTSCs does indeed increase with increasing pressure. To test our findings, we plan to analyze the temperature dependences of PG at different pressures in a $Y_{1-x}Pr_xBa_2Cu_3O_{7-\delta}$ single crystal at $x = 0.57$.

## ACKNOWLEDGMENTS

The work was supported by the National Academy of Sciences of Ukraine within the F19-5 project as well as by the grant of the Finland Academy 367561. One of us (A.L.S.) thanks the Division of Low Temperatures and Superconductivity, INTiBS Wroclaw, Poland, as well as the Physical Division of Lappeenranta Technical University (LUT), Lappeenranta, Finland, for their hospitality.